\documentclass[prc,superscriptaddress,twocolumn]{revtex4}
\usepackage{amssymb,color,graphicx,bm}
\usepackage{amssymb}
\usepackage{amsmath}
\usepackage{graphicx}
\usepackage{hyperref}

\hypersetup{
colorlinks   = true, %Colours links instead of ugly boxes
  urlcolor     = blue, %Colour for external hyperlinks
  linkcolor    = blue, %Colour of internal links
  citecolor   = red %Colour of citations
}

\definecolor{red}{rgb}{0.8,0,0}
\definecolor{RED}{rgb}{0.8,0,0}
\definecolor{violet}{rgb}{0.4,0,0.4}
\definecolor{green}{rgb}{0,0.5,0.0}
\definecolor{GREEN}{rgb}{0,0.5,0.0}
\definecolor{navy}{rgb}{0.0,0.0,0.6}
\definecolor{orange}{rgb}{0.8,0.2,0.0}
\definecolor{blue}{rgb}{0.3,0.0,0.8}

\begin{document}

%\title{Stability and maximum mass of Super-Chandrasekhar white dwarfs with a varying magnetic field}
\title{\bf{Study of motion around a static black hole in  Einstein and Lovelock gravity}}

%Stable equilibrium models of magnetized super-Chandrasekhar white dwarfs with a varying magnetic field

\author{Mukul Bhattacharya$^1$, Naresh Dadhich$^{2,3}$, Banibrata Mukhopadhyay$^1$\\
1. Department of Physics, Indian Institute of Science, 
Bangalore 560012, India\\ 
2. Centre for Theoretical Physics, Jamia Millia Islamia, New Delhi 110025, India\\
3. Inter-University Centre for Astronomy and Astrophysics, Post Bag 4, Pune 411007, India\\
mukul@ug.iisc.in, nkd@iucaa.ernet.in, bm@physics.iisc.ernet.in\\
%\affiliation{Department of Physics, Indian Institute of Science, 
%       Bangalore 560012, India}
%$^*$Corresponding Author
}

\begin{abstract}
We study motion around a static Einstein and pure Lovelock black hole in higher dimensions. It is known that in higher dimensions, bound orbits exist only for pure Lovelock black hole in all even dimensions, $D=2N+2$, where $N$ is degree of  Lovelock polynomial action. In particular, we compute periastron shift and light bending and the latter is given by one of transverse spatial components of Riemann curvature tensor. We also consider pseudo-Newtonian potentials and Kruskal coordinates. \\

%{\it PACS Nos.:} 26.60.-c, 95.30.-k, 87.50.C-, 71.70.Di, 26.60.Kp

\end{abstract}

\maketitle

%\textbf{Keywords}: white and brown dwarfs, magnetic fields, supernova type Ia - standard candles     \\ \\

\section{Introduction}

Recently Dadhich et al. \cite{dadhich} showed that pure Lovelock (the equation of motion has only one term corresponding to a fixed 
$N$)  gravity exhibits bound orbits
and marginally stable circular orbits around a static black hole in all even $D=2N+2 \geq 4$ dimensions, where $N$ is degree of Lovelock polynomial action. This is in contrast to Einstein gravity where bound orbits exist only in $4$ dimensions. Lovelock is the most natural generalization of Einstein gravity and its most remarkable unique property is that inspite of action being polynomial in Riemann curvature, it yields second order equation. It includes Einstein gravity in the linear order $N=1$, Gauss-Bonnet (GB) gravity in the quadratic order $N=2$ and so on. 

It would therefore be of interest to study particle and photon orbits around a static Einstein and pure Lovelock (henceforth Lovelock will stand for pure Lovelock) black hole. That is what would be our concern in this paper. For existence of bound orbits what is required is balance between gravitational and centrifugal potentials. The former becomes stronger with dimension for Einstein gravity as it falls off as $1/r^{D-3}$, while the latter always falls off as $1/r^2$. Thus for existence of bound orbit what is required is $2>D-3$ and hence no bound orbit exists for $ D>4$. On the other hand, Lovelock potential falls off as $1/r^{1/N}$ for even dimensions $D=2N+2$, and  since $1/N<2$ for $N\ge1$, bound orbits will always exist in all even dimensions. Lovelock gravity also exhibits another remarkable property that it is like Einstein in $3$ dimension, kinematic \cite{dadhich1}  relative to $N$th order Riemann and Ricci  curvatures \cite{dadhich2, dadhich3}. In all odd $D=2N+1$ dimensions gravity is kinematic and there exist analogues of BTZ black holes \cite{btz} in all odd dimensions. Lovelock gravity is dynamic in even $D=2N+2$ dimensions and that is why bound orbits exist in all even dimensions.   \\

A static spherically symmetric $D$-dimensional metric is defined as
\begin{equation}
ds^{2} = -f(r)dt^{2} + \frac{1}{f(r)}dr^{2} + r^{2}d\Omega_{D-2}^{2},
\end{equation}
%{\bf{where, $d\Omega_{D-2}^{2}$ is the metric on a $(D - 2)$-dimensional unit sphere and is given by,}}
%\begin{equation}
%\begin{split}
%d\Omega_{D-2}^{2} = d\theta_{1}^{2} + sin^{2}\theta_{1}d\theta_{2}^{2} + sin^{2}\theta_{1}sin^{2}\theta_{2}d\theta_{3}^{2}\\ + \dot{c} + \left[\prod_{j=1}^{D-2}sin^{2}\theta_{j}d\theta_{D-1}^{2} \right]
%\end{split}
%\end{equation}
The effective potential for a test particle moving in such a spherically symmetric space-time is given by 
\begin{equation}
V^{2} = f(r)\left(\frac{l^{2}}{r^{2}} + 1\right). 
\end{equation}
For a static black hole in higher dimensional Einstein gravity \cite{nuovo} we have 
\begin{equation}
f(r) = 1 - \frac{2M}{r^{n}},
\end{equation}
while for Lovelock gravity \cite{cai}
\begin{equation}
f(r) = 1 - \left(\frac{2M}{r}\right)^{1/N},
\end{equation}
where $D=n+3$ and $D=2N+2$, respectively. Here $M$ is an integration constant, which is proportional to the mass ($\mu$) of the black hole and is given by 
\begin{equation}
M = \frac{16\pi G\mu}{(D-2)\Omega_{D-2}},
\end{equation}
where $G$ is the Newton's gravitation constant. In particular for $D=4$, $\Omega_{2}=4\pi$ and $M=2G\mu$.
Clearly it is the function $f(r)$ that governs motion around the black hole. \\

In this paper, we wish to explore motion around a static black hole in Lovelock gravity, and compare and contrast it 
with higher dimensional Einstein gravity. Earlier, geodesic equations in higher dimensional Schwarzschild, Schwarzschild-(anti) de Sitter, Reissner-Nordstr\"om and Reissner-Nordstr\"om-(anti) de Sitter spacetimes were explored in detail \cite{eva}. The paper is organized as follows. In the next section we obtain threshold radii for existence, boundedness and stability for circular orbit which is followed by consideration of periastron shift and light bending. In particular light bending turns out to be proportional to one of spatial components of Riemann curvature tensor. Next we obtain the pseudo-Newtonian potential and alternatively obtain the circular orbits threshold radii for boundedness and stability and energy of the marginally stable circular orbit. We also consider Kruskal extension of these black hole metrics and show that only Lovelock metrics accord to the usual form involving exponentials, while for Einstein in higher dimensions there is an additional algebraic factor. We end with a discussion. 

\vspace{0.2 in}
\section{Circular orbits}
%\vspace{0.1 in}
It is shown in Ref. \cite{dadhich} that for Einstein gravity bound orbits exist in no other dimension than $4$ while for Lovelock gravity they exist in all even dimensions, $D=2N+2$. We recall the discussion of existence, boundedness and stability of circular orbits in Einstein and Lovelock gravities. The conditions for existence of circular orbits are $\dot{r} = \ddot {r} = 0$, bound orbit condition is $E^2=1$ and stability is given by the minimum of effective potential, where $r$ and $E$ are the radius of the orbit and specific energy of the test particle, respectively. Photon circular orbit defines the existence threshold.   

\vspace{0.1 in}
\subsection{Einstein gravity}
For the Einstein solution (3), we have  
\begin{equation}
{E}^{2} = \left(1 - \frac{2M}{r^{n}} \right)\left(1 + \frac{l^{2}}{r^{2}}\right),
\end{equation}
\begin{equation}
\frac{l^{2}}{r^{2}} = \frac{nM}{r^{n}-(n+2)M}.
\end{equation}
From these we obtain the existence, stability and boundedness thresholds as 
\begin{equation}
r_{ph} = \left((n+2)M\right)^{1/n}, 
\end{equation}
%stability threshold 
\begin{equation}
r_{s} = \left(\frac{n+2}{2-n} 2M\right)^{1/n},
\end{equation}
and %and boundedness threshold 
\begin{equation}
r_{b} = \left(\frac{4M}{2-n}\right)^{1/n},
\end{equation}
respectively.
It is clear that bound orbits can exist only for $n=1$ i.e. $D=4$. 
The energy at the marginally stable circular orbit is given by
\begin{equation}
{E_{s}}^{2} = \left(1 - \frac{2M}{r_{s}^{n}} \right)\left(1 + \frac{l^{2}}{r_{s}^{2}}\right) = \frac{8n}{(n+2)^{2}}.
\end{equation}
%Using (7) in (12) we obtain
%\begin{equation}
%{E_{s}}^{2}= \frac{8n}{(n+2)^{2}} .
%\end{equation}
We get back the Schwarzschild values for $n=1$. 

\begin{figure}[h]
\centering
\includegraphics[width=0.45\textwidth]{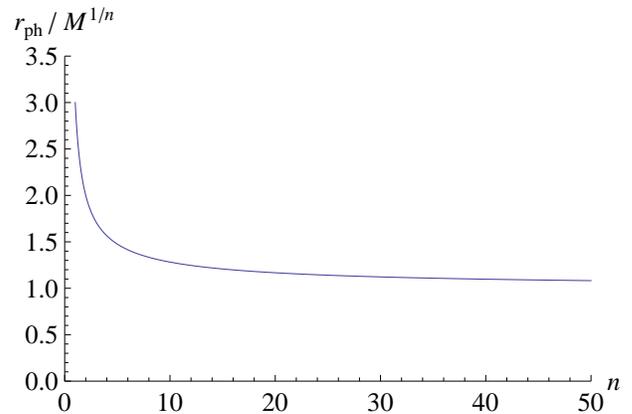}
\caption{\small Variation of $r_{ph}$ as a function of $n$ for Einstein gravity.}
\end{figure}
Figure $1$ shows that the radius of photon orbit gradually decreases with the increase of $n$, because event horizon size also gradually shrinks with increasing $n$.

%\begin{figure}[h]
%\centering
%\includegraphics[width=0.65\textwidth]{einsteinenergystablegr1.pdf}
%\caption{\small Variation of $E_{s}$ as a function of $n$ for Einstein gravity}
%\end{figure}

\vspace{0.1 in}
\subsection{Lovelock gravity}
The corresponding relations for Lovelock solution (4) are given by 
\begin{equation}
{E}^{2} = \left(1 - \left(\frac{2M}{r}\right)^{1/N} \right)\left(1 + \frac{l^{2}}{r^{2}}\right),
\end{equation}
\begin{equation}
\frac{l^{2}}{r^{2}} = \frac{(2M)^{1/N}}{r^{1/N}-(2M)^{1/N}},
\end{equation}
and 
\begin{equation}
r_{ph} = \left(\frac{2N+1}{2N}\right)^{N} 2M,
\end{equation}
\begin{equation}
r_{s} = \left(\frac{2N+1}{2N-1}\right)^{N} 2M,
\end{equation}
\begin{equation}
r_{b} = 2M\left(\frac{2N}{2N-1}\right)^{N}.
\end{equation}
Clearly bound orbits and thereby circular orbits can exist for any $N\ge1$ in all even dimensions, $D=2N+2$. This is a unique property of (pure) Lovelock gravity. 
 
The energy at the marginally stable circular orbit is given by
\begin{equation}
{E_{s}}^{2} = \left(1 - \left(\frac{2M}{r_{s}} \right)^{1/N} \right)\left(1 + \frac{l^{2}}{r_{s}^{2}}\right) = \frac{8N}{(2N+1)^{2}}.
\end{equation}
%From (15)
%\begin{equation}
%{E_{s}}^{2} = \frac{8N}{(2N+1)^{2}, the similarity between $4$ Schwarzschild case and $6$ dimensional pure GB case is quite visible as against $6$ dimensional Einstein case. 

%\end{equation}
For $N=1$, it is Einstein gravity and we get back all Schwarzschild values.  
\begin{figure}[h]
\centering
\includegraphics[width=0.5\textwidth]{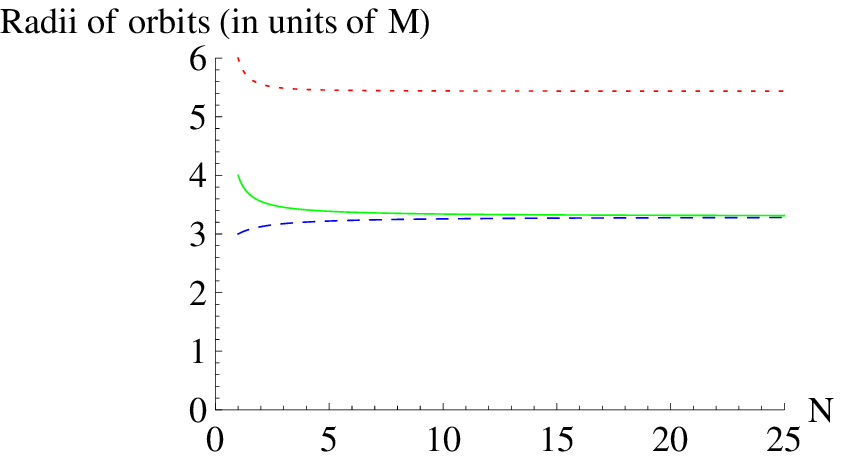}
\caption{\small Variations of $r_{ph}$ (dashed line), $r_{b}$ (solid line) and $r_{s}$ (dotted line)  as functions of $N$ for pure Lovelock gravity.}
\end{figure}

Figure $2$ shows that with the increase of $N$, the strength of the potential increases at a given $r$, making $r_{ph}$ merge with $r_{b}$. Increasing strength of gravitational potential decreases $r_{s}$ and $r_{b}$. This is similar to what is observed in Kerr spacetime as compared to Schwarzschild spacetime. Figure $3$ confirms that with increasing $N$, the particle is more bound at $r_{s}$.

\begin{figure}[h]
\centering
\includegraphics[width=0.45\textwidth]{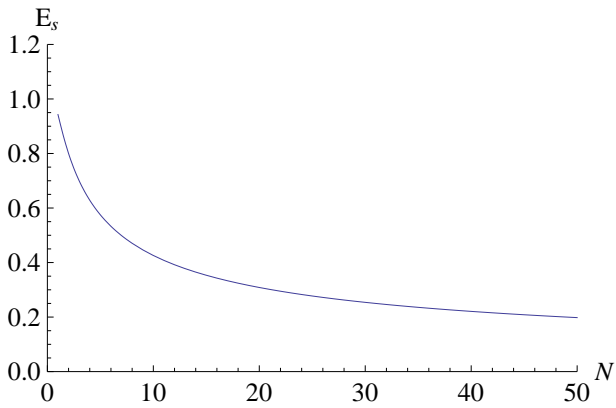}
\caption{\small Variation of $E_{s}$ as a function of $N$ for pure Lovelock gravity.}
\end{figure}

\vspace{0.1 in}
\section{Light bending and space curvature}
%\vspace{0.05 in}
It is clear that light cannot feel $\nabla\Phi$ like ordinary massive particles, hence it can only respond to gravity through spacetime curvature. In here we wish to demonstrate that it is infact proportional to transverse component of Riemann curvature. We follow the standard calculations and use the same notations as used in Ref.  \cite{magnan} for the Schwarzschild case. 
%\vspace{0.05 in}
\subsection{Einstein gravity}
In spherical symmetry, irrespective of additional angular dimensions, there are two constants of motion. 
We define them as 
%For Einstein solution, the Lagrangian is given by 
%the metric is given using $(2)$ and $(4)$ by
%\begin{equation}
%ds^2 = -(1 - \frac{2M}{r^{n}})dt^{2, the similarity between $4$ Schwarzschild case and $6$ dimensional pure GB case is quite visible as against $6$ dimensional Einstein case. 
%} + (1 - \frac{2M}{r^{n}})^{-1}dr^{2} + r^{2}d\Omega^{2}.
%\end{equation}
%\begin{equation}
%L = - \left(1 - \frac{2M}{r^{n}}\right)\dot{t}^{2} + \left(1 - \frac{2M}{r^{n}}\right)^{-1}\dot{r}^{2} + r^{2}\dot{\theta}^{2} + r^{2}sin^{2}\theta\dot{\phi}^{2}.
%\end{equation}
%The constants of motion are the specific energy $E = \left(1 - \frac{2M}{r^{n}}\right)\dot{t}$ and specific angular momentum $l = r^2 \dot{\phi}$ for $\theta=\pi/2$. It then readily follows 
specific energy $E = \left(1 - \frac{2M}{r^{n}}\right)\dot{t}$ and specific angular momentum 
$l = r^2 \dot{\phi}$ (which are the same as those for $D=4$ with $\theta=\pi/2$). 
%{\bf{where $\theta$ and $\phi$ are same as $\theta_{1}$ and $\theta_{2}$ in Eqn. $(2)$.}} 
It then readily follows 
\begin{equation}
\left(\frac{1}{r^{2}} \frac{dr}{d\phi}\right)^{2} = \left(\frac{E}{l}\right)^{2} - \left(1 - \frac{2M}{r^{n}}\right)\left[\left(\frac{m}{l}\right)^{2} + \frac{1}{r^{2}}\right].
\end{equation}
For photon $m=0$ and hence
\begin{equation}
\left(\frac{1}{r^{2}} \frac{dr}{d\phi}\right)^{2} = \left(\frac{E}{l}\right)^{2} - \left(1 - \frac{2M}{r^{n}}\right)\frac{1}{r^{2}}.
\end{equation} 
%The similarity between Schwarzschild case and $6$-dimensional pure GB case is quite visible as against $6$-dimensional Einstein case.

Let $r=R$, represent the point where photon is closest to the source. Writing $l/E = b$ and at $r=R$, $dr/dt=0$, we have
\begin{equation}
\left(\frac{1}{r^{2}} \frac{dr}{d\phi}\right)^{2} = \left(1 - \frac{2M}{R^{n}}\right)\frac{1}{R^{2}} - \left(1 - \frac{2M}{r^{n}}\right)\frac{1}{r^{2}}.
\end{equation}
We substitute $u=R/r$, such that $u$ lies between $0$ and $1$, and obtain
\begin{equation}
d\phi = \frac{\left(1-u^2\right)^{-1/2} du}{\left[1-\frac{2M}{R^{n}}\left(1-u^{n+2}\right)\left(1-u^2\right)^{-1}\right]^{1/2}}.
\end{equation}
We further put $u=cos\alpha$, where $\alpha$ lies between $0$ and $\pi/2$, and use the approximation for small $M/R$ to obtain
\begin{equation}
\phi = \pi + C_{1} \left(\frac{2M}{R^{n}}\right),
\end{equation}
where
\begin{equation}
 C_{1} =  \int^{\pi/2}_{0} d\alpha[(1-{cos}^{n+2}\alpha){sin}^{-2}\alpha].
\end{equation} 
%, the similarity between $4$ Schwarzschild case and $6$ dimensional pure GB case is quite visible as against $6$ dimensional Einstein case. 

Therefore, deflection is given by
\begin{equation}
\Delta\phi = \phi - \pi = C_{1} \left(\frac{2M}{R^{n}}\right).
\end{equation}
We find that the value of $C_{1}$ increases with $n$, as given in Table $1$.

%{\cred GIVE ONLY 4-5 VALUES, 1,2,5,10}

\subsection{Lovelock gravity}
%\vspace{0.05 in}
For Lovelock gravity, the metric is given by $(1)$ and $(4)$. Repeating the same calculation, we obtain  
%\begin{equation}
%ds^2 = -(1 - (\frac{2M}{r}))^{1/N})dt^{2} + (1 - (\frac{2M}{r}))^{1/N})^{-1}dr^{2} + r^{2}d\Omega^{2}.
%\end{equation}
%The corresponding Lagrangian is 
%\begin{equation}
%L = -(1 - (\frac{2M}{r}))^{1/N})\dot{t}^{2} + (1 - (\frac{2M}{r}))^{1/N})^{-1}\dot{r}^{2} + r^{2}\dot{\theta}^{2} + r^{2}sin^{2}\theta\dot{\phi}^{2}.
%\end{equation}
\begin{equation}
\phi = \pi + C_{2} \left(\frac{2M}{R}\right)^{1/N},
\end{equation}
where,
\begin{equation}
C_{2} =  \int^{\pi/2}_{0} d\alpha[(1-{cos}^{2+ 1/N}\alpha){sin}^{-2}\alpha],
\end{equation}
which gives  
\begin{equation} 
\Delta\phi =  C_{2} \left(\frac{2M}{R}\right)^{1/N}.
\end{equation}
The value of $C_{2}$ decreases as $N$ increases and it asymptotically converges to $\pi/2$, as given in Table $1$. 

\vspace{0.1 in}
\subsection{Relation to space curvature}
%\vspace{0.05 in}
Note that light deflection is proportional to gravitational potential, $M/r^n$ and $M/r^{1/N}$ respectively for Einstein and Lovelock black hole. As mentioned before photon can only respond to space curvature, particularly its transverse component, $R^\theta{}_{\phi\theta\phi}$, which is the relevant one for deflection and is in fact given by potential and does not involve its derivative. We can thus say that it is space curvature that bends light and its deflection is proportional to transverse space Riemann component. 

\vspace{0.05 in}
\begin{center}
Table $1$
\end{center}

\begin{center}
Variation of $C_{1}$ and $C_{2}$ with spacetime dimension $D$
\vspace{0.1 in}

\begin{tabular}{|l|l|l|l|l|l|}
\hline
$D$ & $N$ &  $C_{2}$ & $n$ & $C_{1}$ \\ \hline
4 & 1 & 2 & 1 & 2 \\ \hline
6 & 2 & 1.797 & 3 & 2.666 \\ \hline
%3 & 8 & 1.725 & 5 & 8 & 3.12 \\ \hline
10 & 4 & 1.687 & 7 & 3.657 \\ \hline
50 & 24 & 1.591 & 47 & 8.728 \\ \hline
100 & 49 & 1.58 & 97 & 12.439 \\ \hline
%249 & 500 & 1.573 & 497 & 500 & 27.983 \\ \hline
%499 & 1000 & 1.572 & 997 & 1000 & 39.6 \\ \hline
%2499 & 5000 & 1.571 & 4997 & 5000 & 88.61 \\ \hline
%4999 & 10000 & 1.571 & 9997 & 10000 & 125.32 \\ \hline
\end{tabular}
\end{center}

\vspace{0.2 in}
\section{Bound orbits}
%\vspace{0.05 in}
As shown in the Introduction, for Einstein gravity bound orbits exist only in $4$ dimensions and none else, while for Lovelock they exist in all even dimensions, $D=2N+2$. We shall consider the two cases of bound orbits corresponding to Lovelock $N=1, 2$, that would include usual Schwarzschild case for $N=1$ and $D=4$ and GB case for $N=2$ and $D=6$. We would contrast it with Einstein in $6$ dimensions for $n=3$. 

The orbit equation in the usual notation for Einstein in $4$ and $6$ dimensions is given by 
\begin{equation}
u^{"} + u = \frac{M}{l^{2}} + 3Mu^{2},
\end{equation}
\begin{equation}
u^{"} + u = \frac{3Mu^{2}}{ l^{2}} + 5Mu^{4},
\end{equation}
respectively, while for GB it is 
\begin{equation}
u^{"} + u = \frac{(2M)^{1/2}u^{-1/2}}{4 l^{2}} + \frac{5(2M)^{1/2}u^{3/2}}{4}.
\end{equation}

\begin{figure}[h]
\centering
\includegraphics[width=0.5\textwidth]{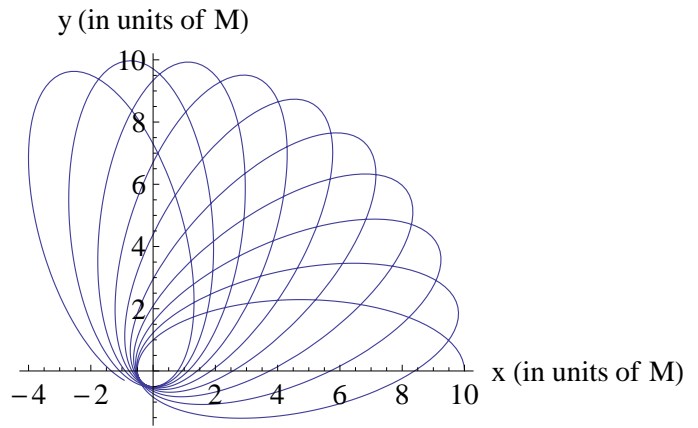}
\caption{\small Cartesian plot for Schwarzschild case: $u^{"} + u = 1 + 0.03u^{2}$, where $u=1/r$, using $M=0.01$ and $l=0.1$ and boundary conditions $u(0)=0.1$ and $u^{'}(0)=0$}
\end{figure}

\begin{figure}[h]
\centering
\includegraphics[width=0.5\textwidth]{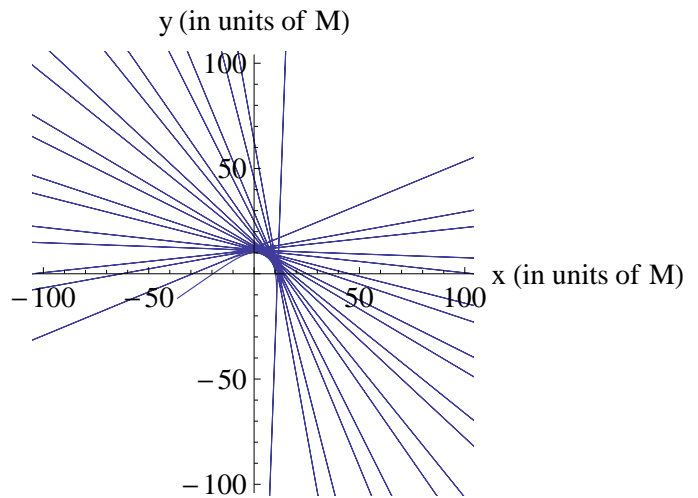}
\caption{\small Cartesian plot for Einstein $6D$ case:   $u^{"} + u = 3 u^{2} + 0.05 u^{4}$, where $u=1/r$, using $M=0.01$ and $l=0.1$ and boundary conditions $u(0)=0.1$ and $u^{'}(0)=0$ }
\end{figure}

\begin{figure}[h]
\centering
\includegraphics[width=0.5\textwidth]{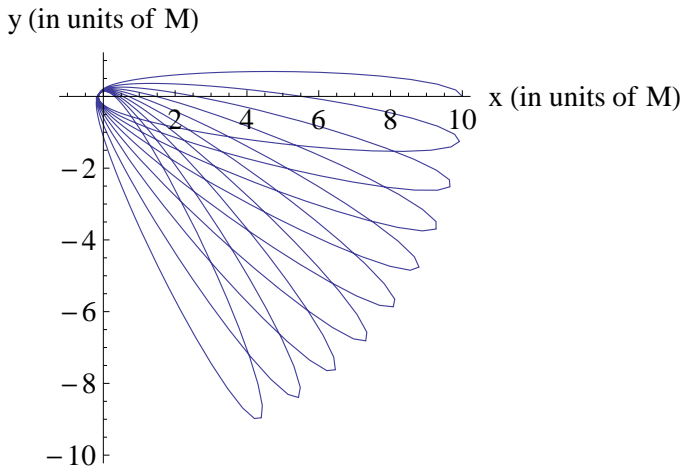}
\caption{\small Cartesian plot for Pure Lovelock $6D$ case:  $u^{"} + u = 3.535 u^{-1/2} + 0.17675 u^{3/2}$, where $u=1/r$,  using $M=0.01$ and $l=0.1$ and boundary conditions $u(0)=0.1$ and $u^{'}(0)=0$ }
\end{figure}
Equation $(28)$ can be solved analytically with some approximations and we obtain a bound orbit which is elliptical in shape with precessing periastron, and it is as given in any textbook, for example, \cite{weinberg}. 
However in the other two cases, it is quite difficult to solve the differential equations analytically and we instead solve them numerically. 

For Schwarzschild case, we solve $u^{"} + u = 1 + 0.03u^{2}$  by using $M=0.01$ and $l=0.1$. We use the boundary conditions, $u(0)=0.1$ and $u^{'}(0)=0$. By solving this numerically we obtain elliptical orbits with periastron shifts as expected, shown in Fig. $4$.

For Einstein $6 D$ case, $u^{"} + u = 3 u^{2} + 0.05 u^{4}$, using $M=0.01$ and $l=0.1$. For boundary conditions $u(0)=0.1$ and $u^{'}(0)=0$, we see that there does not exist any bound orbit as shown in Fig. $5$. This is consistent with the fact that bound orbits cannot exist in Einstein gravity unless $D=4$ \cite{dadhich}.

For Lovelock $6 D$ case, $u^{"} + u = 3.535 u^{-1/2} + 0.17675 u^{3/2}$, using $M=0.01$ and $l=0.1$. For boundary conditions $u(0)=0.1$ and $u^{'}(0)=0$, we do have bound orbits with precession as exhibited in Fig. $6$. However the orbit is not exactly elliptical and hence shift cannot be called ``periastron shift". Here we have shown that bound orbits exist for $N=2$, similarly they would do so for all $N$ in $D=2N+2$. 

\vspace{0.2 in}
\section{Pseudo-Newtonian potential}
The concept of pseudo-Newtonian potential is useful in understanding fluid flow around black holes. Ref. \cite{m02} introduces a methodology to evaluate pseudo-Newtonian potential for any metric in general. In the present context, we would like to use it for an alternative computation of circular orbit threshold radii and interestingly it turns out that we obtain the same values as obtained earlier, using general relativity. Note that centrifugal force would go as $1/r^3$, while the gradient of pseudo potential would give gravitational force.  

Pseudo-Newtonian potential \cite{m02} for Einstein and Lovelock cases are given by 
\begin{equation}
V = - \frac{M}{r^{n}-2M},
\end{equation}
and 
\begin{equation}
V = - \frac{(2M)^{1/N}}{2(r^{1/N}-(2M)^{1/N})},
\end{equation}
respectively. Then corresponding forces would respectively be given as 
\begin{equation}
F = nM\frac{r^{n-1}}{(r^{n}-2M)^{2}},
\end{equation}
and
\begin{equation}
F = \frac{(2M)^{1/N}}{2N}\frac{r^{1/N -1}}{(r^{1/N}-(2M)^{1/N})^{2}}.
\end{equation}

The Newtonian energy, which is free of rest mass, is defined as 
\begin{equation}
\Tilde{E} = \frac{v^{2}}{2} + V = \frac{r}{2}\frac{dV}{dr} + V = \frac{r}{2} F + V.
\end{equation}
The bound orbit threshold radius would be given by $\Tilde{E}=0$, while stability threshold would be given by $dl/dr=0$, and we obtain 
\begin{equation}
r_{b} = \left(\frac{4M}{2-n}\right)^{1/n},
\end{equation}
and
\begin{equation}
r_{s} = \left(\frac{n+2}{2-n}2M\right)^{1/n},
\end{equation}
for the Einstein case, while for Lovelock case, 
%\begin{equation}
%r_{s} = 2M\left(\frac{2N+1}{2N-1}\right)^{N},
%\end{equation}
\begin{equation}
r_{b} = 2M\left(\frac{2N}{2N-1}\right)^{N},
\end{equation}
and 
\begin{equation}
r_{s} = 2M\left(\frac{2N+1}{2N-1}\right)^{N}.
\end{equation}
\vspace{0.05 in} 
These are indeed the same as obtained earlier in Sec II. 

%{\cred THE DISCUSSIN OF ENERGY IS NOT VERY ILLUMINATING AND HENCE IT COULD BE DROPPED.} 

\begin{figure}[h]
\centering
\includegraphics[width=0.45\textwidth]{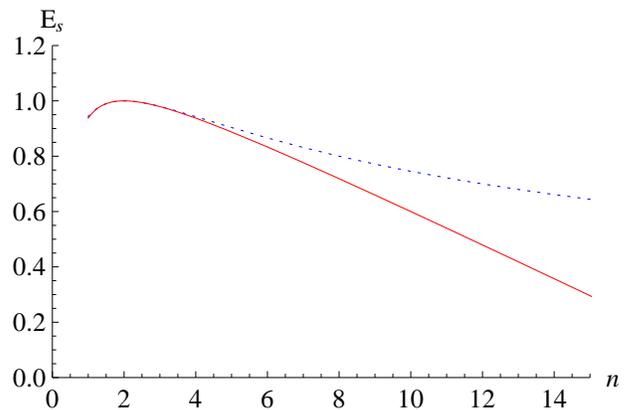}
\caption{\small Variation of energy of marginally stable circular orbit for Einstein (dotted line) and pseudo-Newtonian gravity (solid line) with $n$}
\end{figure}

%\subsection{Energy of marginally stable orbit}
%Energy of last stable circular orbit from General relativity, 
%\begin{equation}
%\epsilon_{s}=\frac{(8N)^{0.5}}{2N+1}
%\end{equation}
The energy of marginally stable circular orbit using pseudo-Newtonian potential is given by
\begin{equation}
E_{s}=1 + \tilde{E_{s}}=1 - \frac{(2N-1)^{2}}{16N}.
\end{equation}
\begin{figure}[h]
\centering
\includegraphics[width=0.45\textwidth]{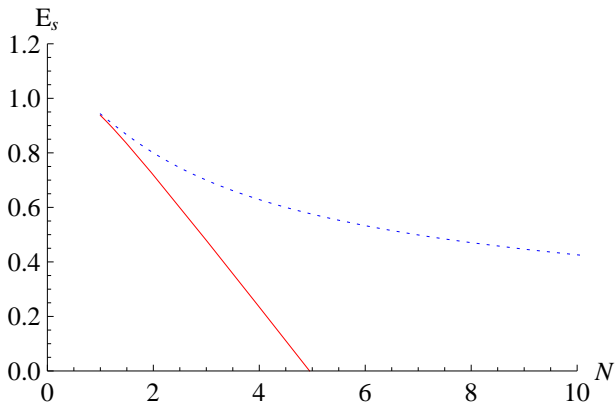}
\caption{\small Variation of energy of marginally stable circular orbit for pure Lovelock (dotted line) and pseudo-Newtonian (solid line) gravity with $N$}
\end{figure}
From Figs. $7$ and $8$, we see that the energy of the marginally stable circular orbit as predicted by the pseudo-Newtonian potential is very close to that of general relativity for the Schwarzschild case i.e. $n=N=1$, but is not so for the higher dimensions. The difference between energies of the last stable circular orbit increases with spacetime dimension for both Einstein and pure Lovelock gravities.

\vspace{0.2 in}
\section{Kruskal coordinates}
We can carry out Kruskal coordinate transformation for higher dimensional Einstein and Lovelock solutions following the same procedure as that for Schwarzschild metric. We shall perform Kruskal transformation for $ D= 4, 5$ for Einstein and $D=6, 8$ for Lovelock black hole.  

\subsection{Schwarzschild case ($D=4, n=N=1$)}
First we recapitulate the transformation for Schwarzschild metric. We look at null geodesics with $d\theta = d\phi = 0$ and use $(3)$ for $n=1$ to obtain
\begin{equation}
dt = \pm \frac{dr}{\left(1 - \frac{2M}{r}\right)}.
\end{equation}
Integrating $(41)$, we obtain $t = \pm r^{*}+constant$, where $r^{*}$ is the tortoise coordinate given by
\begin{equation}
\small r^{*}=r+2M log\left(\frac{r}{2M} - 1\right).
\end{equation}
The metric then becomes 
\begin{equation}
ds^{2}=\left(1-\frac{2M}{r}\right)\left(-dt^{2} + dr^{*2}\right) + r^{2} d\Omega_{2}^{2}.
\end{equation}
We further define Eddington-Finkelstein coordinates as $\tilde{u}=t+r^{*}$ and $\tilde{v}=t-r^{*}$. In terms of $\tilde{u}$ and $\tilde{v}$, the metric takes the form 
\begin{eqnarray}
ds^{2} = \frac{1}{2}\left(1-\frac{2M}{r}\right)\left(d\tilde{u}d\tilde{v} + d\tilde{v}d\tilde{u}\right) + r^{2}d\Omega_{2}^{2}.
\end{eqnarray}
Still $r=2M$ corresponds to either $\tilde{u}=-\infty$ or $\tilde{v}=\infty$, so we further transform coordinates to pull these points to finite coordinate values  as
\begin{equation}
u^{'}=e^{\tilde{u}/4M},  \hspace{0.2 in}
v^{'}=e^{-\tilde{v}/4M}.
\end{equation}
We thereafter define $u=\frac{1}{2}(u^{'}-v^{'})$ and $v=\frac{1}{2}(u^{'}+v^{'})$. The metric finally comes to the desired form 
\begin{equation}
ds^{2}=\frac{32 M^3}{r}e^{-r/2M}(du^{2}-dv^{2})+r^{2}d\Omega_{2}^{2}.
\end{equation}
The coordinates ($v,u,\theta,\phi$) are the Kruskal coordinates in which metric is regular everywhere except at centre $r=0$, which is the curvature singularity. For constant $r$, $u^2 - v^2 = constant$, which is a hyperbola in $u-v$ plane.

\vspace{0.1 in}
\subsection{Einstein gravity ($n=2,3; D= n+3$)}
\vspace{0.1 in}
\subsubsection{\bf{Einstein 5D $(n=2)$}}
Using $(3)$ for $n=2$, we write
\begin{equation}
dt = \pm \frac{dr}{\left(1 - \frac{2M}{r^{2}}\right)},
\end{equation}
which gives $t = \pm r^{*}+constant$, where
\begin{equation}
r^{*}=r+\frac{(2M)^{1/2}}{2}log\left(\frac{r-\sqrt{2M}}{r+\sqrt{2M}}\right)
\end{equation}
and the metric reads 
\begin{equation}
ds^{2}=\left(1-\frac{2M}{r^{2}}\right)\left(-dt^{2} + dr^{*2}\right) + r^{2} d\Omega_{3}^{2}.
\end{equation}
Further to Eddington-Finkelstein coordinates,  
\begin{equation}
ds^{2} = \frac{1}{2}\left(1-\frac{2M}{r^{2}}\right)\left(d\tilde{u}d\tilde{v} + d\tilde{v}d\tilde{u}\right) + r^{2}d\Omega_{3}^{2}.
\end{equation}
We then transform the coordinates to pull points $\tilde{u}$ and $\tilde{v}$ corresponding to $r=(2M)^{1/2}$ to finite coordinate values as
\begin{equation}
u^{'}=e^{\tilde{u}/(2M)^{1/2}},  \hspace{0.2 in}
v^{'}=e^{-\tilde{v}/(2M)^{1/2}},
\end{equation}
and finally we obtain  
\begin{equation}
\small ds^{2}=\frac{4 M^{2}}{r^{2}} e^{-2r/\sqrt{2M}}\left(\frac{r}{\sqrt{2M}}+1\right)^{2}(-dv^{2}+du^{2})\\+r^{2}d\Omega_{3}^{2}.
\end{equation}
%It is still singular at $r=2GM$, and we have verified that it is so for $D=6$ as well. This indicates that Kruskal like coordinates in which metric is free of coordinate singularity may not exist for Einstein gravity in $D>4$. 
%\vspace{0.1 in}

%{\cred i presume there is typo, r/\sqrt{2GM} +1, it should be -1, pl. do check it.} IF THAT IS SO THEN WE NEED NOT GIVE D=6 CASE, THE ABOVE STATEMENT IS ENOUGH} - Actually it is not a typo, it comes out to be +1 and not -1. Kruskal coordinates are defined in a way for Schwarzschild, Einstein and Lovelock gravities such that there exists only a singularity at r=0, singularity at r=2GM or (2GM)^(1/2) or (2GM)^(1/3) do not exist for any of the Kruskal coordinates, either Einstein or Lovelock.

\subsubsection{\bf{Einstein 6D $(n=3)$}}
For $n=3$, we obtain
\begin{equation}
dt = \pm \frac{dr}{\left(1 - \frac{2M}{r^{3}}\right)}.
\end{equation}
As before, $t = \pm r^{*}+constant$, where %$r^{*}$ is the tortoise coordinate and is given by
\begin{equation}\small
\begin{split}
r^{*}=r+(2M)^{1/3}[\frac{1}{3}log\left(\frac{r}{(2M)^{1/3}}-1\right)\\ - \frac{1}{6}log\left(\frac{r^{2}+(2M)^{1/3}r+(2M)^{2/3}}{(2M)^{2/3}}\right)\\-\frac{1}{\sqrt{3}}{\tan}^{-1}\left(\frac{2r+(2M)^{1/3}}{\sqrt{3}(2M)^{1/3}}\right)].
\end{split}
\end{equation}
%Writing the metric in terms of tortoise coordinate
%\begin{equation}
%ds^{2}=(1-\frac{2GM}{r^{3}})(-dt^{2} + dr^{*2}) + r^{2} d\Omega^{2}
%\end{equation}
%We define Eddington-Finkelstein coordinates as $\tilde{u}=t+r^{*}$ and $\tilde{v}=t-r^{*}$. In terms of $\tilde{u}$ and $\tilde{v}$ we can write the metric as
%\begin{equation}
%ds^{2} = \frac{1}{2}(1-\frac{2GM}{r^{3}})(d\tilde{u}d\tilde{v} + d\tilde{v}d\tilde{u}) + r^{2}d\Omega^{2}
%\end{equation}
The forms of the metric in terms of tortoise and Eddington-Finkelstein coordinates are the same as $(49)$ and $(50)$ except for $2 M/r^{2}$ replaced by $2 M/r^{3}$.
We transform the coordinates to pull $\tilde{u}$ and $\tilde{v}$ corresponding to $r=(2M)^{1/3}$ to finite coordinate values as
\begin{equation}
u^{'}=e^{3\tilde{u}/2(2M)^{1/3}},  \hspace{0.2 in}
v^{'}=e^{-3\tilde{v}/2(2M)^{1/3}}.
\end{equation}
We then define $u=\frac{1}{2}(u^{'}-v^{'})$ and $v=\frac{1}{2}(u^{'}+v^{'})$. The metric in terms of $u$ and $v$ is
\begin{eqnarray}
\begin{split}
\tiny ds^{2}=\frac{8}{9}\frac{(2M)^{5/3}}{r^{3}}e^{-3r/(2M)^{1/3}} \\ \left(\frac{r^{2}+(2M)^{1/3} r+(2M)^{2/3}}{(2M)^{2/3}}\right)^{3/2} \\e^{{\sqrt{3}\tan^{-1} (\frac{2r+(2M)^{1/3}}{\sqrt{3}(2M)^{1/3}}})} (du^{2}-dv^{2})+r^{2}d\Omega_{4}^{2}.
\end{split}
\end{eqnarray}\\

The coordinate system ($v,u,\theta,\phi$) form the Kruskal coordinate system where $v$ is the time-like coordinate.

\vspace{0.1 in}
\subsection{Lovelock $(N=2, 3, D=2N+2)$}
We have already considered $N=1$ case for Schwarzschild solution in $D=4$. We would now consider the cases $N=2, 3$ and show that proper Kruskal coordinates exist in which metric is free of coordinate singularity. 

\subsubsection{\bf{Gauss-Bonnet $(N=2, D=6)$}}
Using $(4)$ for $N=2$, we obtain
\begin{equation}
dt = \pm \frac{dr}{\left(1 - (\frac{2M}{r})^{1/2}\right)},
\end{equation}
which gives $t = \pm r^{*}+constant$ , where
\begin{equation}
r^{*}=r+(2M)^{1/2}[2\sqrt{r}+2\sqrt{2M}log\left(\frac{\sqrt{r}-\sqrt{2M}}{\sqrt{2M}}\right)].
\end{equation}
We follow the same steps as before and first to tortoise coordinates
\begin{equation}
ds^{2}=\left(1-\left(\frac{2M}{r}\right)^{1/2}\right)(-dt^{2} + dr^{*2}) + r^{2} d\Omega_{4}^{2},
\end{equation}
and then to Eddington-Finkelstein coordinates 
\begin{equation}
ds^{2} = \frac{1}{2}\left(1-\left(\frac{2M}{r}\right)^{1/2}\right)\left(d\tilde{u}d\tilde{v} + d\tilde{v}d\tilde{u}\right) + r^{2}d\Omega_{4}^{2}.
\end{equation}

Similarly, writing as before
\begin{equation}
u^{'}=e^{\tilde{u}/8M},  \hspace{0.2 in}
v^{'}=e^{-\tilde{v}/8M}.
\end{equation}
and defining $u=\frac{1}{2}(u^{'}-v^{'})$ and $v=\frac{1}{2}(u^{'}+v^{'})$, we obtain the metric in Kruskal coordinates 
\begin{equation}
ds^{2}=\frac{32(2M)^{5/2}}{r^{1/2}}e^{-r/4M}e^{-\sqrt{r/2M}}(-dv^{2}+du^{2})+r^{2}d\Omega_{4}^{2}.
\end{equation}
%This is indeed regular as desired everywhere except at $r=0$. 

\vspace{0.1 in}
\subsubsection{\bf{Lovelock $(N=3, D=8)$}}
Using $(4)$ for $N=3$ gives
\begin{equation}
dt = \pm \frac{dr}{(1 - \left(\frac{2M}{r}\right)^{1/3})},
\end{equation}
which gives $t = \pm r^{*}+constant$, where 
\begin{equation}
\begin{split}
r^{*}=r+\frac{3}{2}(2M)^{1/3}r^{2/3}+6 Mlog\left(\left(\frac{r}{2M}\right)^{1/3}-1\right) \\+3(2M)^{2/3}r^{1/3}.
\end{split}
\end{equation}
%We can write the metric in terms of tortoise coordinate as
%\begin{equation}
%ds^{2}=(1-(\frac{2GM}{r})^{1/3})(-dt^{2} + dr^{*2}) + r^{2} d\Omega^{2}
%\end{equation}
%In terms of Eddington Finkelstein coordinates, the metric is
%\begin{equation}
%ds^{2} = \frac{1}{2}(1-(\frac{2GM}{r})^{1/3})(d\tilde{u}d\tilde{v} + d\tilde{v}d\tilde{u}) + r^{2}d\Omega^{2}
%\end{equation}
The forms of the metric in terms of tortoise and Eddington-Finkelstein coordinates are the same as $(59)$ and $(60)$ except $(2M/r)^{1/2}$ is replaced by $(2M/r)^{1/3}$.
We transform the coordinates to pull the points $\tilde{u}$ and $\tilde{v}$ corresponding to $r=2M$ to finite coordinate values as 
\begin{equation}
u^{'}=e^{\tilde{u}/12M},  \hspace{0.2 in}
v^{'}=e^{-\tilde{v}/12M}.
\end{equation}
The metric in terms of $u$ and $v$ (using the same definition as earlier) is given by 
\begin{equation}
\begin{split}
ds^{2}=\frac{72(2M)^{7/3}}{r^{1/3}}e^{-r/6M}e^{-\frac{1}{2}(r/2M)^{2/3}}e^{-(r/M)^{1/3}}\\(-dv^{2}+du^{2})+r^{2}d\Omega_{6}^{2}.
\end{split}
\end{equation}
%This is also regular everwhere except at $r=0$. \\

As expected, the potential retains its $r^{-k}$ dependence in all the cases. Lovelock case is distinguished from Einstein in that the metric in Kruskal extension always involves exponentials as against additional algebraic factor for Einstein in higher dimensions. 

%The important conclusion that emerges is that Kruskal extension works for Lovelock gravity in all even $D=2N+2$ dimensions which includes Einstein Schwarzschild solution for $N=1$ in $D=4$. In contrast, for higher dimensional Einstein solutions, following the standard procedure we have not been able to find proper Kruskal extension in which metric is regular everywhere except at the centre. \\

%\section*{Acknowledgment} 
%The work has been supported by the project with Grant No. ISRO/RES/2/367/10-11.
%The authors would like to thank the referee for useful comments and 
%suggestions to improve the quality of the paper.

\vspace{0.2 in}
\section{Discussion}
It is argued in Refs \cite{dadhich1, dadhich3} that Lovelock gravity has similar behaviour in odd $(2N+1)$ and even $(2N+2)$ dimensions; i.e. similar to Einstein gravity for $N=1$ in $3$ and $4$ dimensions. Here we have employed particle orbits around static black hole and its Kruskal extension to establish this feature. For instance, existence of bound orbits is a universal common feature for Lovelock gravity in all even dimensions. It is also interesting to see that light bending is proportional to transverse spatial Riemann component,  $R^{\theta}_{\phi\theta\phi}$. This is to indicate that light does not experience acceleration $\nabla\Phi$, but simply follows the curvature of space. Light bending is in reality space bending which is measured by means of light. For Lovelock black hole, there is always a proper Kruskal extension in all even $D=2N+2$ dimensions in which the metric is free of coordinate singularity similar to $4$-dimensional Schwarzschild solution. In contrast, Einstein metrics in higher dimensions have additional algebraic factor. \\

Our main aim has been to probe the universal common behaviour of Lovelock gravity by the 
study of motion around a static black hole, and it has been fully borne out. However one may ask the question, 
how does the higher dimensional gravitational dynamics affect our observational $4$-dimensional Universe. 
For instance, in the Kaluza-Klein $5$-dimensional theory, an additional dimension incorporates electromagnetic 
field. That is, to get higher dimensional effects onto the usual spacetime, some additional prescription 
has to be imposed like the Kaluza-Klein compactification of extra dimension. For the case of pure Lovelock, 
the immediate next higher order is quadratic GB, which is topological in $D=4$. It could however 
be made non-trivial by coupling it with a scalar field --- dilaton. There have been extensive studies 
of dilaton field \cite{brax, cho}, however they all refer to Einstein-GB. In line with the viewpoint 
advocated in this paper and elsewhere \cite{dadhich3, dadhich1, dadhich}, it should be pure GB coupled 
to a scalar field without the Einstein-Hilbert $R$ for probing higher dimensional effects. Very recently, 
a very interesting and novel inflationary model has been found \cite{kgd} with a scalar field coupled 
to pure GB Lagrangian. Another very popular method is brane world gravity \cite{randal-sundrum} in which 
curvature of higher dimensional bulk spacetime is employed to make extra dimension small enough. May what 
all that be, this has not been our concern in this paper.  

%\vspace{0.2 in}
\section*{Acknowledgement}
%\vspace{-0.18 in}
MB and ND thank Prof. Lars Andersson for warm hospitality at AEI, Potsdam-Golm, where large part of the work was done and MB also thanks DAAD (German Academic Exchange Service) for the summer visiting fellowship.

\end{document}